\documentclass[aps,pre,reprint,groupedaddress]{revtex4-1}
\usepackage{graphicx}
\usepackage{dcolumn}
\usepackage{bm}
\usepackage{amsmath}
\usepackage{amsfonts}
\usepackage{subeqnar}
\usepackage{color}

\begin{document}


\title{The role of multistability and transient trajectories in networked dynamical systems:  
Connectomic dynamics of \emph{C. elegans} and behavioral assays}

\author{James Kunert$^1$\footnote{Electronic address: \texttt{kunert@uw.edu}}, Eli Shlizerman$^2$, Andrew Walker$^2$ and J. Nathan Kutz$^2$}%
%
\affiliation{$^1$ Department of Physics, University of Washington, Seattle, WA. 98195}

\affiliation{$^2$ Department of Applied Mathematics, University of Washington, Seattle, WA. 98195-2420}

\date{\today}

\begin{abstract}
The neural dynamics of the nematode~\emph{C. elegans} are experimentally low-dimensional and correspond to discrete behavioral states, where previous modeling work has found neural proxies for some of these states. Experimental results further suggest that dynamics may be understood as long-timescale transitions between multiple low-dimensional attractors. To identify multistable regimes of our model, we develop a method for systematic generation of bifurcation diagrams and their analysis in an interpretable low-dimensional subspace, showing the existence and nature of multistable input responses at a glance. Stimulation of the PLM neuron pair, experimentally associated with forward movement and shown in simulation to drive a limit cycle, defines our low-dimensional projection space. We then obtain bifurcation diagrams for single-neuron excitation over a range of amplitudes and which classify whether the dynamics in this projection space are associated with a limit cycle, fixed point, or multiple states. In the specific case of compound input into both the PLM pair and ASK pair we discover bistability of a limit cycle and a fixed point, with transitional timescales between different states being much longer than other timescales in the system. This suggests consistency of our model with the characterization of dynamics in neural systems as long-timescale transitions between discrete, low-dimensional attractors corresponding to behavioral states. Our methodology thus prescribes a method for identifying these states and transitions in response to arbitrary input.
\end{abstract}


\maketitle
\section{Introduction}
Understanding the functional responses and control of high-dimensional networked dynamical systems is of critical importance across the physical, engineering and biological sciences.   In many such systems, even with large numbers of nonlinearly interacting nodes, meaningful input and output are dominated by low dimensional spatio-temporal patterns of activity~\cite{holmes,Kutz:2013,Mezic11}.  Indeed, the underlying networked dynamics can be thought to be dominated by trajectories that evolve on low-dimensional attractors and/or induced transient trajectories between attractors. As a specific biophysical example, neuronal networks, which are typified by high-dimensional networks of neurons, display robust functional responses and behavioral assays that are encoded by such low-dimensional attractors or transient trajectories~\cite{jones07,rabinovich01,rabinovich08,rabinovich11,laurent01,gold99,eli,Kunert}.

The nematode~\emph{Caenorhabditis elegans} is an important model system in understanding how these neuronal networks generate robust functional responses to inputs, partly due to the fact that the connectivity between its 302 neurons (its connectome) has been resolved~\cite{White,Varshney}. \emph{C. elegans} is capable of a wide range of behaviors over various timescales~\cite{Faumont}, yet experimental studies suggest that these behaviors are fundamentally low-dimensional~\cite{Bialek08}, and the behaviors of the worm can be understood as low-dimensional trajectories on attractors between which it will transition stochastically~\cite{Bialek11}.

While the exact role of the connectome in neuronal computation remains unresolved and in general controversial, it has been shown that simple models of \emph{C. elegans} neural dynamics (combining specific connectivity data with simple unfit parameter estimates and dynamics) are capable of generating non-trivial, qualitatively correct responses to given stimuli~\cite{Kunert}. This suggests that such computational modeling can be informative in understanding how the system generates behavioral responses. It is therefore of interest to consider whether or not models that are capable of producing neural proxies for behavioral responses (as in~\cite{Kunert}) are further capable of characterizing experimentally observed multistable attractor dynamics.

One motive in the search for multistability is that the previous study in~\cite{Kunert} finds a neural proxy for behavior consisting simply of a single limit cycle within the system. On the other hand, experimental evidence suggests that many neural responses are better described as transient trajectories between multiple attractors~\cite{rabinovich11}, rather than the traditional dynamical systems view in which behavior is described by dynamics on a stable attractor. Within this paradigm it is important not only that multistability should exist, but that transients with biophysically relevant behavioral timescales should exist.

In this paper, we explore the input space of our full model for the neuronal network dynamics of \emph{C. elegans}, developed in~\cite{Kunert}, and find that various multistabilities arise in response to inputs. Performing direct neuronal simulations to reveal such transitions is a formidable task, since the input space is large and neuronal simulations produce high-dimensional outputs which are difficult to interpret. We therefore develop a systematic methodology to explore responses to complex inputs and understand the dynamics within a framework of low-dimensional attractor dynamics. Specifically, for a chosen input vector we generate a bifurcation diagram (using the amplitude of the constant-in-time input as our bifurcation parameter) to identify multistability. We use an interpretable low-dimensional projection (as defined by forward motion) to characterize the dynamics between multiple attractors as identified by the bifurcation diagrams.

With this framework we survey all input vectors corresponding to single-neuron current injections and find bifurcations corresponding to limit cycles and multiple attractors. For some of these input vectors which induce multistability, simulated transient dynamics are on much slower (on the order of seconds to tens of seconds) than any intrinsic neuronal timescales (which in our model do not exceed a few hundred milliseconds). The transient trajectories themselves are low-dimensional and could be associated with network-produced functionalities, such as neural proxies for movement.  Thus our simulations of connectomic dynamics are in agreement with behavioral observations of {\em C. elegans} and help support recent biophysical conjectures that the transients themselves are critical in understanding behavioral assays~\cite{rabinovich11}.

As a particular example, we choose input into the PLM neuron pair, which is known experimentally to excite forward motion~\cite{Chalfie} and within our model creates a two-dimensional limit cycle response~\cite{Kunert}. We then use the low-dimensional PLM response plane to consider the dynamics of a compound input vector PLM+ASK, where ASK stimulation is known to facilitate transitions (i.e. turns)~\cite{Gray04}. Our bifurcation analysis reveals that this induces bi-stability, in which the system goes either into a fixed point or a limit cycle. Transient timescales are shown to be considerably longer in this bistable case than the intrinsic timescales of the system. This allows for long timescales in the system arising from transitions between discrete, low-dimensional attractors corresponding to behavioral states, consistent with the experimentally-based framework~\cite{Bialek11}. This input scenario demonstrates how our bifurcation analysis methodology prescribes a generic approach for identifying multi-stable states and transitions between them in response to arbitrary inputs. Since we model neurons as identical except for their connectivity, it further indicates that their connectivity alone can encode the creation and destruction of multiple behavioral attractors and transitions between them.

\section{Simulation of \emph{C. elegans} Connectome Dynamics}

\begin{figure}
\includegraphics[width=1.0\columnwidth]{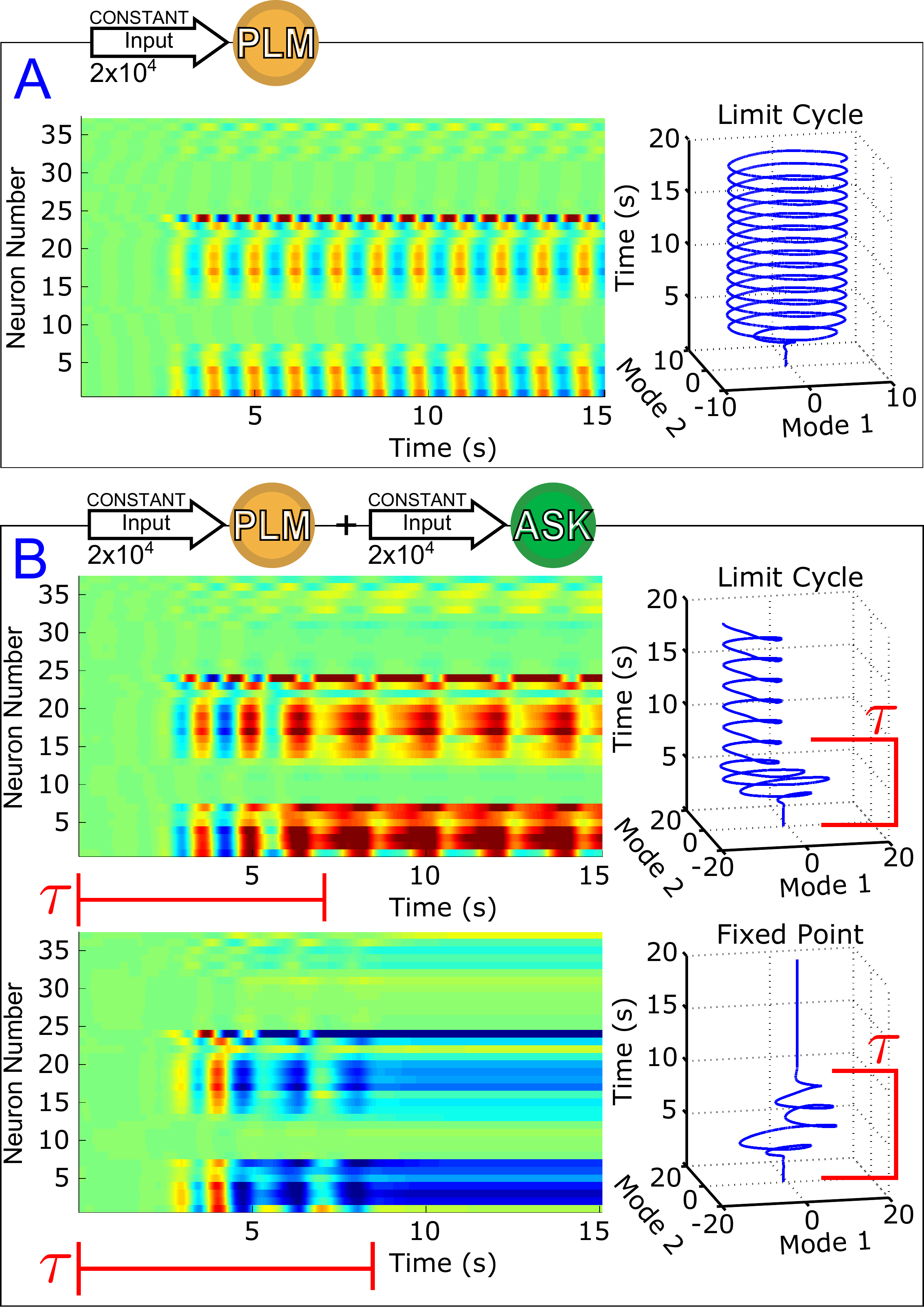}
\caption{\label{fig1}
Voltage dynamics of forward-motion motorneurons (neurons of classes DB, DD, VB and VD) in response to the following sensory inputs: in Panel A, an input of $2\times 10^4$ (Arb. Units) into the PLM sensory neuron pair (known experimentally to drive forward motion~\cite{Chalfie}); in Panel B, an input of $2\times 10^4$ into the PLM pair with an additional input of $2.5\times 10^4$ into the ASK sensory neuron pair (known experimentally to promote turning~\cite{Gray04}). Simultaneous PLM+ASK stimulation causes bistability, with relatively long transient times $\tau$. To the right of each raster plot is the trajectory within the Forward-Motion 2D Plane (defined by the trajectory in Panel A, and used for all subsequent projections).
}
\end{figure}

\subsection{Model for Coupled Neural Dynamics}
The dynamic model used here is constructed to represent the graded responses of the neurons of \emph{C. elegans}. Experiments show that many neurons in the organism are effectively isopotential, so that the membrane voltage can be used as a state variable for characterizing the neuron~\cite{Goodman98}. Wicks et al.~\cite{Wicks96} used this to construct a single-compartment membrane model for neuron dynamics. Building on these findings in~\cite{Kunert}, we were able to construct a full connectomic dynamics model in which it was shown to yield reasonable low-dimensional neural proxies for known behavioral responses (specifically, it was shown that simulating excitation of the tail-touch mechanosensory pair PLM creates a two-mode oscillatory limit cycle in the forward motion motorneurons). As in~\cite{Kunert}, neural membrane voltage dynamics are governed by:
\begin{equation}\label{eq:dvdt}
C\dot{V_i} = -G^{c} (V_i-E_{cell})-I_i^{Gap}(\vec{\textbf{V}})-I_i^{Syn}(\vec{\textbf{V}})+I_i^{Ext}
\end{equation}
$C$ is the whole-cell membrane capacitance, $G^{c}$ is the membrane leakage conductance  and  $E_{cell}$ is the leakage potential. The external input current (which we change to specify the external stimulus) is given by $I_i^{Ext}$, while neural interaction via gap junctions and synapses is modeled by input currents  $I_i^{Gap}(\vec{\textbf{V}})
$ (gap) and $I_i^{Syn}(\vec{\textbf{V}})$ (synaptic).  Their equations
are:
\begin{align}\label{eq:gaps1}
&I_i^{Gap} = \sum\limits_j G_{ij}^g(V_i-V_j)\\
&I_i^{Syn} = \sum\limits_j G_{ij}^{s} s_j(V_i-E_j)
\end{align}
Gap junctions are taken as ohmic resistances connecting each neuron where $G_{ij}^g$ is the total conductivity of the gap junctions between $i$ and $j$. Synaptic current is proportional to the displacement from reversal potentials $E_j$. $G_{ij}^s$ is the maximum total conductivity of synapses to $i$ from $j$, modulated by the synaptic activity variable $s_i$, which is governed by:
\begin{equation}\label{eq:synapticvar}
\dot{s_i} = a_r\phi(v_i;\kappa,V_{th})(1-s_i)-a_ds_i
\end{equation}
where $a_r$ and $a_d$ correspond to the synaptic activity's rise and decay time, and $\phi$ is the sigmoid function $\phi(v_i;\kappa,V_{th})=1/(1+\exp({-\beta(V_i-V_{th}})))$.

We keep the parameters values of~\cite{Kunert}. In particular, the connectivity parameters $G_{ij}^g$ and $G_{ij}^s$ are prescribed by the full connectome~\cite{Varshney}. The relative significance of these specific connectivity values is maintained by not fitting any of the other global parameters. Instead, these parameters are estimated to a reasonable order of magnitude from the literature and assumed equal for each neuron. Relevant values to this section are, as taken from~\cite{Kunert}: gap junctions and synapses are both given individual conductances of $g=100\textrm{pS}$; cell membranes are set to a conductance of $G^c = 10\textrm{pS}$; membrane capacitances are set to $1\textrm{pF}$; and the synaptic rise and decay constants are set to $a_r=\textrm{1 s}^{-1}$ and $a_d=\textrm{5 s}^{-1}$.  Thus all neurons are modeled as identical except for their connectivity and the assignment of them as excitatory or inhibitory (where $E_j$ will have one of two values corresponding to these classes). 

\subsection{\label{ss:time}Model Timescales}

\begin{table}
\begin{tabular}{| l | c |}
\hline
Interaction & Timescale \\
\hline
Single-Neuron Membrane Leakage & 100ms \\
Gap Junctions & 10ms \\
Synaptic Connections & 200ms\\
\hline
\end{tabular}
\caption{\label{tab:times}
Orders of magnitude for various timescales within the system for the parameters chosen.
}
\end{table}

Of particular relevance to this paper are the timescales within the system. From the first term in Equation (\ref{eq:dvdt}), we see that the exponential free decay constant of an unconnected neuron (i.e. decay through the membrane leakage term alone, with $I_i^{Gap}=I_i^{Syn}=I_i^{ext}=0$) would be $\tau_{free} = C/G^c = \textrm{100ms}$. Similarly, the time constant value given by gap junctions would be $\tau_{gap}=C/g = \textrm{10ms}$.

There are also timescales intrinsic to the synaptic dynamics. We approximate these by considering the dynamics when voltages are held constant, and thus $\phi(v_i;\kappa,V_{th}) \equiv \phi_i$ is constant. Then Equation (~\ref{eq:synapticvar}) becomes:
\begin{equation}
\dot{s_i} = a_r \phi_i - (a_r\phi_i + a_d)s_i
\end{equation}
and thus the synapses will exponentially approach equilibrium with a time constant of $\tau_{syn} = 1/(a_r\phi_i+a_d)$. Since $a_r=\textrm{1 s}^{-1}$, $a_d=\textrm{5 s}^{-1}$, and $\phi_i \in (0,1)$, synapses must have exponential time constants in the range $\tau_{syn} \in (166,200) \textrm{ms}$.

It will be shown that, when the system is in a bistable regime, the timescales of transient dynamics within the system can be orders of magnitude above any of these intrinsic time constants within the system (on the order of $10\textrm{s}$, for example).

\subsection{Model Discussion}

The model does not include various extra-synaptic features known to drive or regulate responses. For example, there is evidence that self-sustained forward locomotion in~\emph{C. elegans} is regulated by proprioception within motor neurons~\cite{Aravi12} (compare how our model, lacking this, does not sustain oscillation in the absence of explicit external input). Computational modeling which includes stretch-receptive proprioception shows that such feedback loops can control behavioral features such as gait modulation between differing environments~\cite{Boyle,Bryden}. The lack of such feedback mechanisms and other signaling mechanisms (such as various neuromodulators, monoamines and peptides~\cite{Qi,Vidal}), in combination with the simple neuron model and parameter assumptions, mean that specific responses to given inputs seen within the model can be encoded only within the network's connectivity. This reductive approach yields information as to how behavioral responses could be encoded within the structure of the connectome.

As an example of the model's ability to generate proxies for behavioral responses encoded within the network, it was shown that stimulating the tail-touch mechanosensory neuron pair PLM, which experimentally leads to forward motion~\cite{Chalfie}, gives rise via a bifurcation to a limit cycle within the forward-motion motorneurons. This limit cycle consists of only two modes, in agreement with the behavioral observation that the worm's body shape during forward motion is well-described by a similar two-mode oscillation~\cite{Bialek08}. The non-triviality of this agreement was established by showing that simulated ablation studies affected this response in agreement with experimental ablation studies (e.g., ablation of the AVB interneurons destroys the response both experimentally and in the model~\cite{Chalfie}).

\subsection{\label{secF1}Response of Model to Given Inputs}

Figure~\ref{fig1} shows the response of forward motion motorneurons to various inputs as a function of time. Panel A of the figure shows a raster plot of motorneuron voltages in response to PLM input (through the $I_{ext}$ term in Equation~(\ref{eq:dvdt})), for which the two-mode oscillatory response can be observed~\cite{Kunert}. The trajectory of these two leading modes are plotted as a function of time on the right. We use this same low-dimensional space (defined as the two forward-motion motorneuron modes which oscillate during PLM activation) throughout the paper. In other words, we use the same projection for the low-dimensional trajectories in Panel B and in all further figures.

The response to PLM stimulation alone consists of a single possible state (i.e. a limit cycle trajectory), but if the model is capable of describing the dynamics in terms of long-timescale transitions between states under the same input, then we wish to find inputs which allow multiple states and transitional dynamics. We find that such inputs indeed exist. As an example, we consider the response to simultaneous stimulation of the PLM neuron pair along with the 
ASK  neuron pair. We choose this stimulation since excitation of ASK neurons have been shown experimentally to promote turning~\cite{Gray04} and their ablation greatly increases the duration of periods of forward motion~\cite{Wakabayashi}.

As we show in Figure~\ref{fig1}B, for this combined input there coexist two different attractors, i.e. the system is bistable. The two trajectories plotted are in response to the same constant input amplitudes into PLM and ASK, and differ only by their initial conditions. Note that the transients before convergence into the eventual fixed point or limit cycle have long timescales (relative to the intrinsic timescales of the system as discussed in Section~\ref{ss:time}). The model therefore does exhibit multistability for this given input, but given the large dimensionality of the input space the discovery, identification and interpretation of these multistable regimes is not trivial. Since we wish to understand the neural dynamics as consisting of long-timescale transitions between discrete attractors, we develop a method for (1) identifying the existence and nature of attractors in response to arbitrary inputs, (2) characterizing transient timescales, and (3) providing interpretable biophysical meaning to calculated trajectories via projection onto a meaningful low-dimensional space.

\section{Bifurcation Diagrams for State Identifications}

\subsection{Calculating Bifurcation Diagrams}

\begin{figure}
\includegraphics[width=1.0\columnwidth]{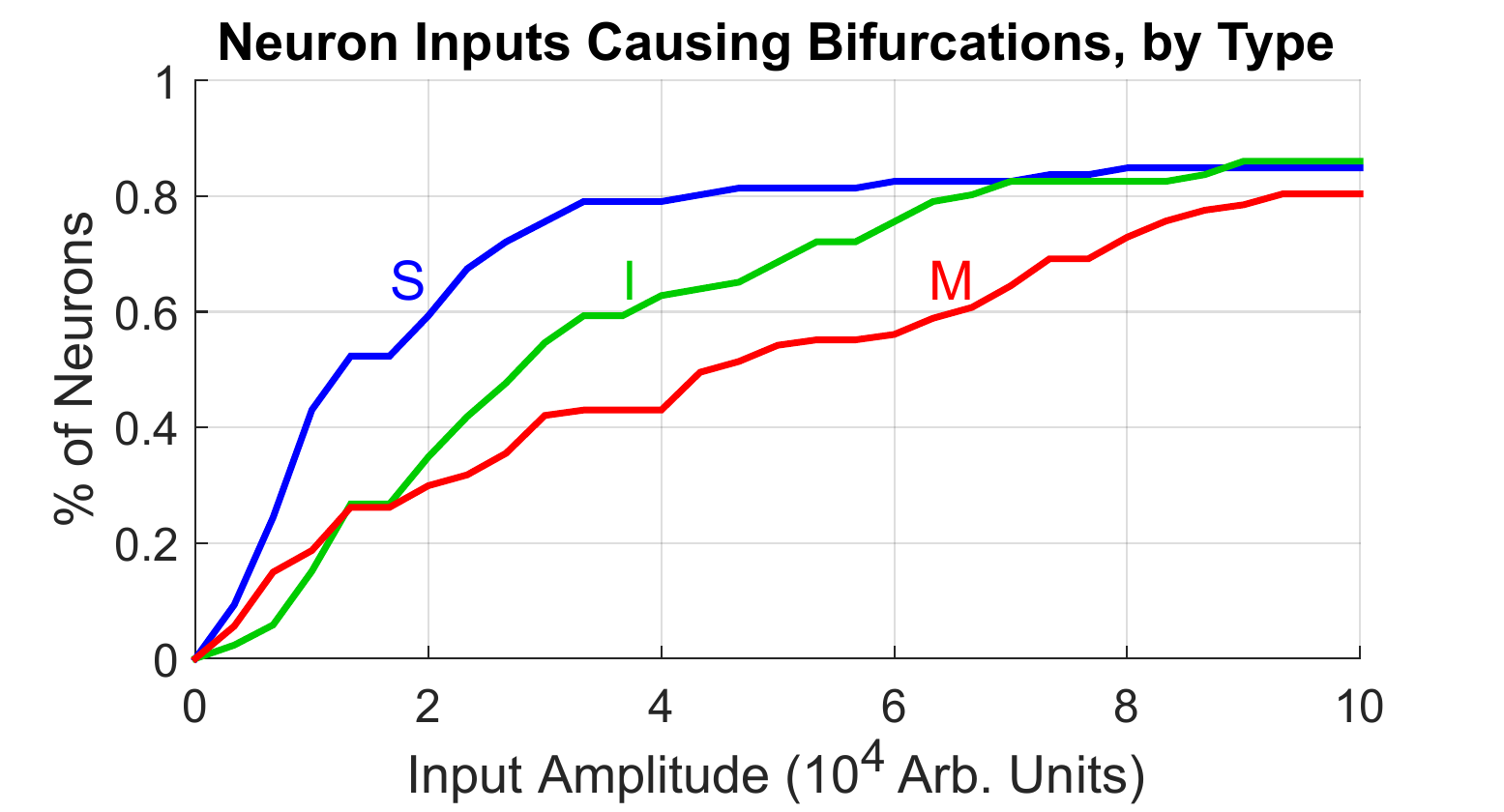}
\caption{\label{fig2a}
The input amplitude into a single neuron at which the standard equilibrium becomes unstable varies by neuron type. The vertical axis shows the percentage of neurons of a given type (sensory neurons, interneurons, or motorneurons) for which the standard equilibrium is unstable when said neurons receive the corresponding input amplitude. Note that sensory neurons , on average, create a bifurcation at a lower input amplitude than do interneurons, which in turn require a lower amplitude than motorneurons.
}
\end{figure}

Motivated by observational studies which describe~\emph{C. elegans} behavioral dynamics in terms of low-dimensional attractor dynamics~\cite{Bialek11}, we wish to understand our simulated neural dynamics as consisting of transitions between discrete attractors. We therefore propose to construct bifurcation diagrams that depict attractors existing within the system under arbitrary inputs. By fixing the direction of the input vector $I_{ext}$ in Eq.~(\ref{eq:dvdt}) and using its amplitude as our bifurcation parameter, such diagrams will show us at a glance the set of states created in response to a given input, and provide us with a method of identifying induced multistability. 

Figures~\ref{fig2} and~\ref{fig3} show examples of such bifurcation diagrams, in which we plot the furthest $L^2$ distance from standard equilibrium (within the 2D Forward-Motion Plane) of all attractors present as a function of input amplitude.  In principle, such diagrams could be calculated by simply performing a large ensemble of simulations and projecting the results into the 2D plane, but such simulations are relatively time-consuming. We can take advantage of the fact that, for this model, it is easy to compute its Jacobian matrix at any given point for any constant input. We therefore use Newton's method when possible, supplementing with simulations to explore the input space and find addtitional attractors or when Newton's method does not converge to the desired result. Full detail on the algorithm used to generate such diagrams can be found in the Supplementary Materials.

We generated these diagrams for all 279 of the single-neuron inputs into the system. Note that the figures from these simulations, as shown in the supplementary materials, are done over a much coarser range than those in Figures~\ref{fig2} and~\ref{fig3}. The purpose of these coarse figures is to quickly give an indication of the likely number of states for each range of inputs. Thus these diagrams give a means of identifying what attractors will exist within the system for a broad range of arbitrary inputs, and of easily identifying regions of multistability in the input space.

Generating these diagrams for all possible single inputs allows for the qualitative comparison of features within each neuron's bifurcation diagram. Similar features in the bifurcation diagrams of neurons may suggest similar functionalities. As a simple example, in Figure~\ref{fig2a}, we compare the input amplitude at which the standard equilibrium first becomes unstable for sensory neurons, interneurons and motorneurons. The majority of sensory neurons are seen to drive bifurcations in the system at lower input levels than for most interneurons, which in turn require lower inputs than most motorneurons. Intuitively, this suggests that the system is typically more sensitive to input into sensory neurons than it is to interneuron or motorneuron inputs. Furthermore, for each group of neurons we compute the percentage of single neuron inputs which promote limit cycle attractors. We find that within our input range, 32.6\% of sensory neurons and 26.7\% of interneurons give rise to oscillatory dynamics, whereas only 8.4\% of motorneurons result in oscillation when stimulted. This points to the sensitivity and particular ability of sensory neurons to drive complex dynamics within the network. Such results serve as a demonstration of the ability of these bifurcation diagrams to provide meaningful and intuitive information about the functionality of neurons and the behavior of the system.

\subsection{A Defining Example: Response to PLM Input}
\begin{figure}
\includegraphics[width=1.0\columnwidth]{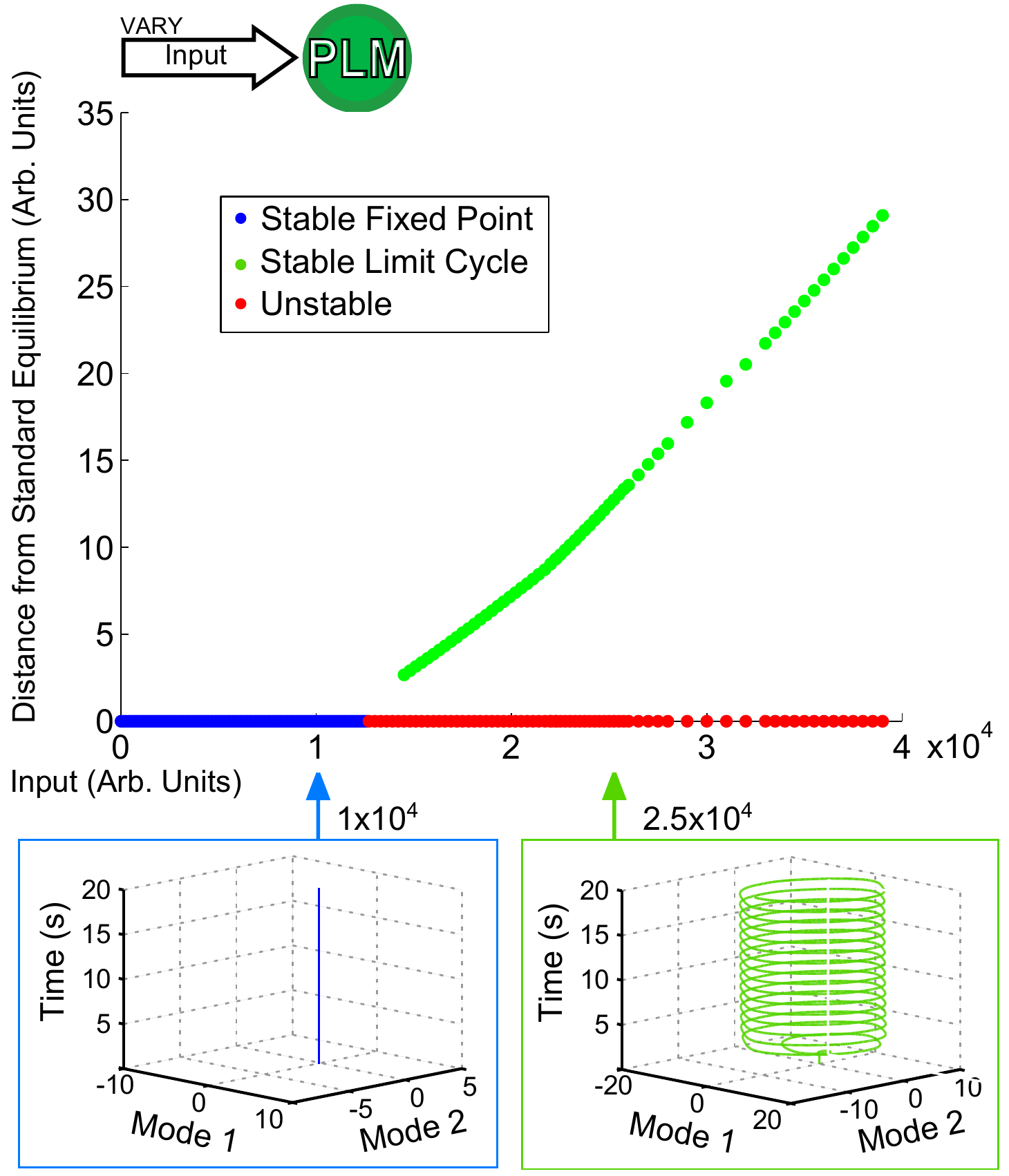}
\caption{\label{fig2}
Bifurcation diagram for constant PLM stimulation of varying amplitude. Below an input of $1.2\times 10^4$ the system goes to a stable fixed point very close to the standard equilibrium, but beyond that input level the system goes to a stable limit cycle (where the plotted point gives the furthest distance from standard equilibrium on the limit cycle). The diagram shows the two qualitatively distinct regions of interest for PLM inputs: the low input level in which the system remains at a fixed point, and the higher input level beyond which the system enters into a limit cycle (which in this case can be considered to serve as a proxy for forward motion~\cite{Kunert}).
}
\end{figure}

Figure~\ref{fig2} shows a low-dimensional bifurcation diagram for constant PLM input. The figure shows the creation of a stable limit cycle in response to input into the neurons PLML/R. By evaluating this bifurcation diagram we can identify the regions of interest which have qualitatively distinct responses (in this case, the region with a lone attractor which is a stable fixed point and the second region with a lone stable attractor which is a limit cycle after the fixed point attractor becomes unstable). For each region we can perform simulations which are then projected onto the low-dimensional plane (the PLM limit cycle being what defines this plane). Given the correspondence of this limit cycle to forward motion, as in~\cite{Kunert}, these low-dimensional trajectories are readily interpretable: the fixed point corresponds to a static worm, and the limit cycle corresponds to oscillatory motion of the body of the worm.

\subsection{Characterizing Bistable Dynamics}

\begin{figure}
\includegraphics[width=1.0\columnwidth]{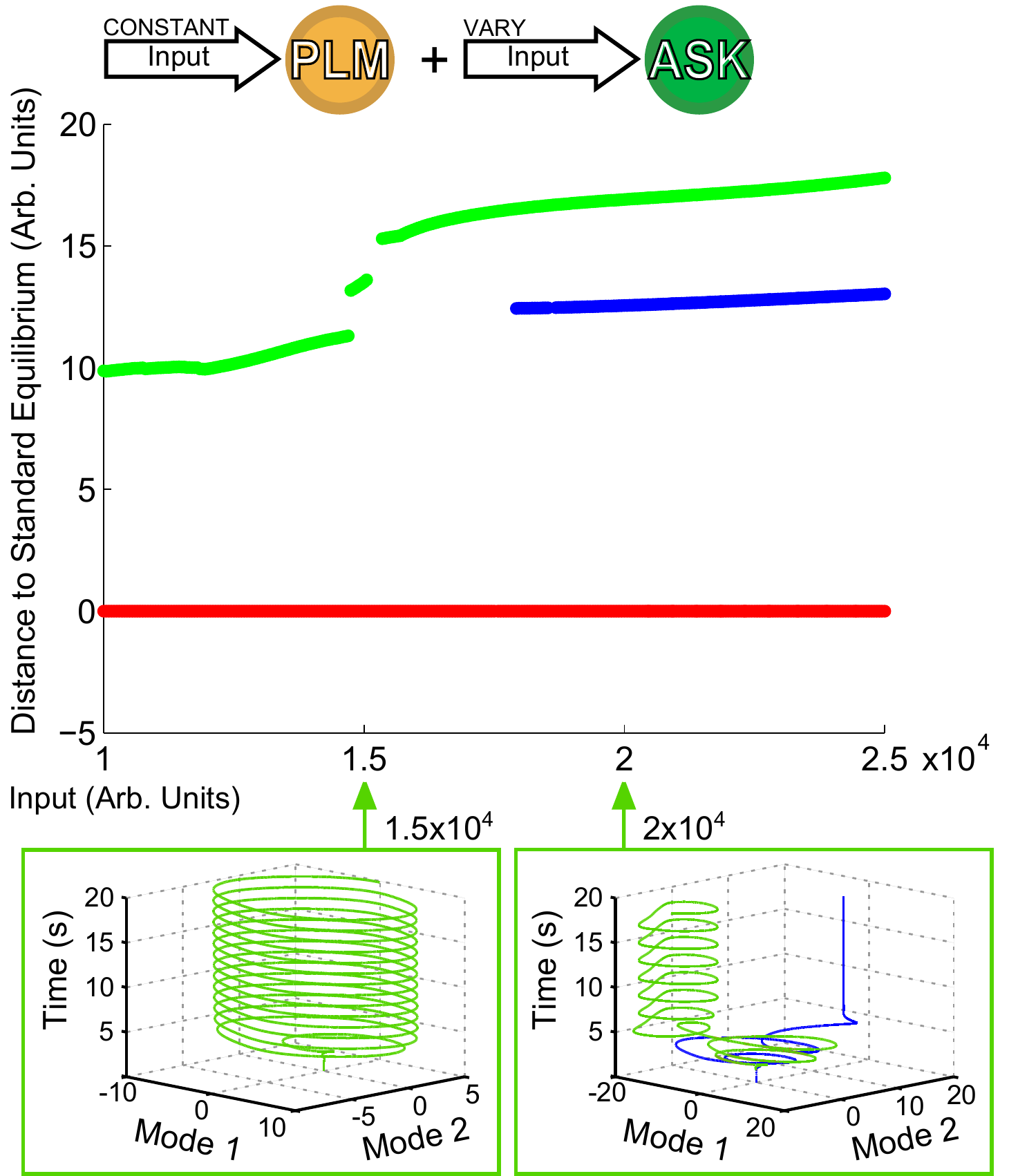}
\caption{\label{fig3}
Bifurcation Diagram for varying amplitude of input into the ASK pair. Input into PLM is fixed at $2\times 10^4$. Note that as input into ASK increases, the forward-motion limit cycle remains relatively undisturbed until it reaches about $1.5 \times 10^4$, after which the distance jumps and a fixed point becomes stable, giving rise to a bistability within the system.
}
\end{figure}

Of greater interest are responses to compound activations; that is, more complicated inputs leading to more complicated responses. We consider as an example the dual input into the PLM and ASK neuron pairs as discussed in Section~\ref{secF1}. We keep a constant input of $2\times 10^4$ into the PLM pair and use as our bifurcation parameter the input into the ASK pair. Figure~\ref{fig3} shows the resulting bifurcation diagram. At inputs below $1.5\times 10^4$, the limit cycle remains relatively undisturbed. At greater inputs, however, a series of bifurcations occur such that there is a sudden jump in the distance of the limit cycle, and at about $1.7\times 10^4$ the system becomes bistable with the addition of a new fixed point. Thus we are able to immediately identify from this figure multistability within the system, which we may then go on to investigate further. Specifically, we are interested in the further investigation of transient timescales of the system.

\section{Long Timescales and Interstate Transients}

\begin{figure}[t]
\includegraphics[width=1.0\columnwidth]{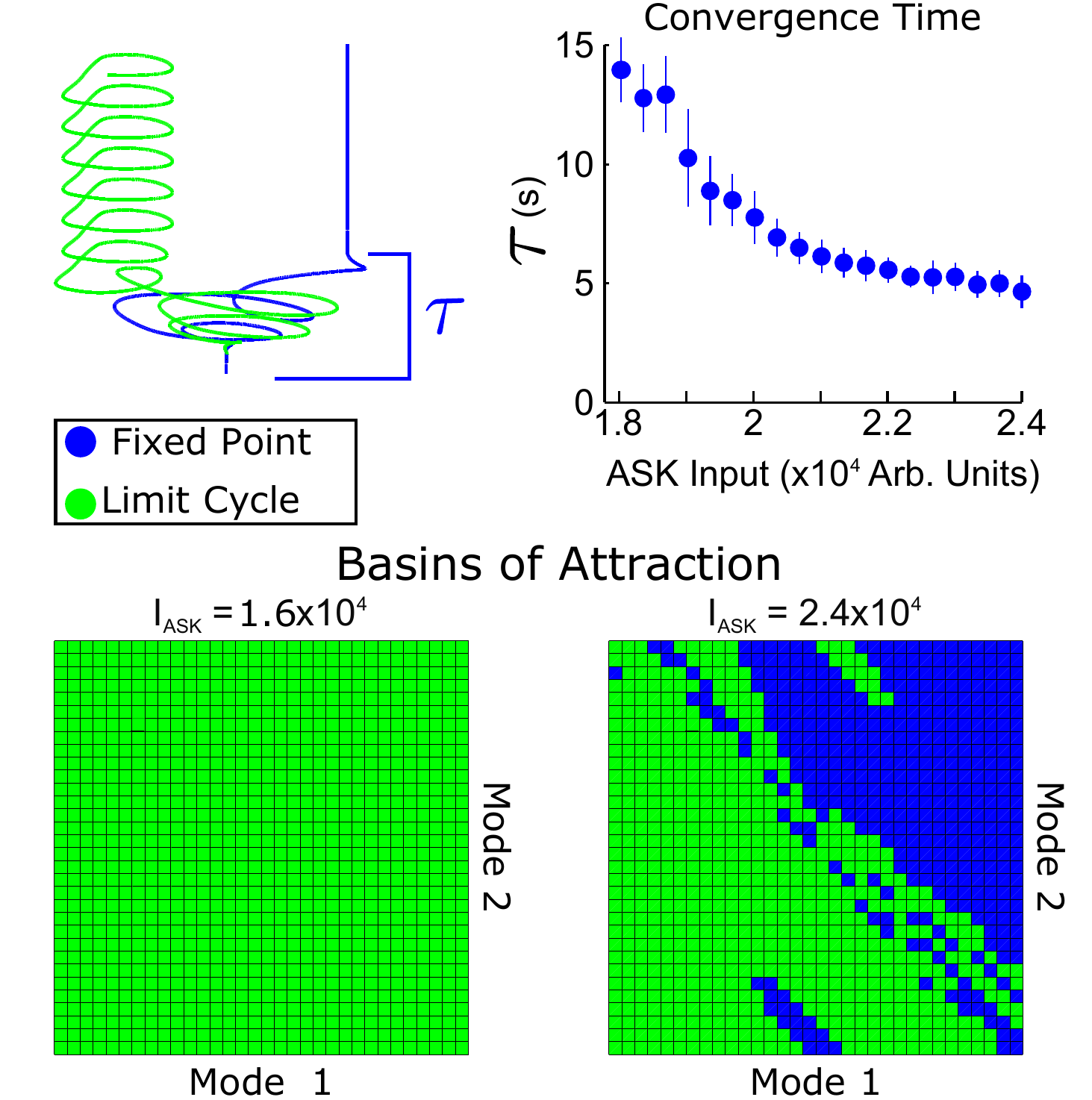}
\caption{\label{fig4}
Spatial and temporal properties of convergence for PLM+ASK input (i.e. the bistable region of Figure~\ref{fig3}). The upper-right plot shows fixed point convergence times as a function of input amplitude (from 200 trials at each point). Note the relatively long transient timescale. The second row shows the spatial basins of attraction for different inputs. Each grid covers a small region around the standard equilibrium, plotting on $(-4,4)\times 10^{-6}$ for both modes. At an ASK input of $1.6\times 10^4$ all initial conditions converge to a limit cycle, but initial conditions on the plane are split between the limit cycle and fixed points at higher inputs such as $2.4\times 10^4$.
}
\end{figure}

In Figure~\ref{fig4} we investigate spatial and temporal aspects of the convergence onto one of the two bistable states. An ensemble of 200 simulations (with random initial conditions in the neighborhood of the standard equilibrium) were performed for each ASK input level. From those, the trials converging to the fixed point solution were taken and the convergence time $\tau$ was calculated by calculating, for each fixed point trial, the time after which all points of the trajectory are within a distance $\epsilon$ of the final value (using here $\epsilon = 0.004$). The average and standard deviation of these convergence times are shown in the top right of Figure~\ref{fig4}. Convergence times for the limit cycle solutions are qualitatively similar when comparing trajectories such as those in the upper-left of the figure. Note that these convergence times are considerably longer than other timescales within the system (comparing, for example, the model's free neuron decay constant of 10ms~\cite{Varshney}, or other trajectory timescales such as the limit cycle period, which remains approximately two seconds regardless of ASK input).

Shown also are the basins of attraction for trajectories starting on the low-dimensional plane, on a grid of initial conditions centered at the standard equilibrium point (which we choose as our origin). The size of the grid is chosen to be within a small neighborhood of zero (within the range $(-4,4)\times10^{-6}$) since we find that trajectories initiated farther away are first attracted towards the zero point before being rerouted to the fixed point or limit cycle attractors. We observe consistency between the structure of these basins of attraction and the observed convergence times. In addition, the basin of attraction plots indicate distinct regions in which initial conditions are more probable to be attracted to the fixed point, and thus shows which portions of forward movement (i.e. which segment of the PLM-driven limit cycle trajectory) are more prone to ASK-driven transitions.

\section{Conclusion}

We explored the input space of a {\em C. elegans} neural dynamic model which incorporates its fully-resolved connectome and demonstrated that various multistabilities arise in response to inputs.  Using a low-dimensional projection space based upon forward motion, we are able to 
systematically explore responses to complex inputs and understand them in a framework of low-dimensional attractor dynamics.  In our study, the bifurcation diagram is constructed by using the constant-in-time input as our bifurcation parameter. We show that such diagrams are capable of revealing and mapping multiple attractors within the system by using a low-dimensional projection space which guides the search for attractors, identifying their stability and their effect upon forward movement.
Furthermore, the low-dimensional projection helps in the interpretation of the dynamics upon the discovered attractors, especially the dynamics associated with multistability. We characterize such multistable dynamics, noting specifically that when the system enters into a multistable regime, transient timescales within the system can be very long relative to intrinsic neural timescales (comparing, for example, the three orders of magnitude between the $\sim$100ms neural timescales in Section \ref{ss:time} to the $\sim$10s transient lengths in Figure \ref{fig4}).

The fact that multistability within the connectomic dynamical system is capable of generating such long transient timescales has critical implications. These longer timescales, on the order of many seconds, are on a similar order to many behavioral timescales such as forward crawling survival time~\cite{Bialek11}. This suggests that various behavioral responses could be associated not with the attractor itself, but rather with the transient leading to that attractor. This is consistent with theoretical constructions and experimental observations of transient orbits between attractors~\cite{rabinovich11,rabinovich01,rabinovich08}.  Importantly, this viewpoint is supported {\em independently} and in a completely different theoretical framework by direct connectomic simulations from biophysically appropriate neuron dynamics within the worm, i.e. the multistability of attractors and long-time transients are not engineered in the model to fit the data and observations, rather they naturally arise from the dynamics associated with the connectome.

This study suggests that neural computations can consist of both dynamics on attractors (as in our PLM-driven limit cycle) and of long-timescale transients between multiple attractors which may arise in the system (as we show in the long-timescale transients between the multistable states from PLM+ASK input). We have demonstrated that both dynamical features can arise by applying simple, identical neuron models onto the \emph{C. elegans} connectome data, suggesting that these responses are encoded within the connectome itself. 

More broadly, many networked dynamical systems across the engineering, physical and biological sciences may also be
dominated by patterns of activity and long-time transients induced by the structure of the network architecture.  Understanding
the basic principles of such behaviors is critical for optimizing performance and controlling deleterious effects.  One can
easily imagine scenarios in which suppressing transients would be important, such as the observed power-grid swing instability~\cite{Mezic11}.  The analysis above may be able to help understand how the network architecture encodes such deleterious
patterns of activity when combined with relevant dynamics.  In contrast, one might desire to generate a network architecture to induce a transient that is beneficial for some purpose relative to an application (for instance, a crawling motion in the case of the {\em C. elegans}).  Understanding how the network connectivity graph drives such activity would be critical for inducing such beneficial patterns of activity, perhaps even suggesting network control protocols for achieving desired results.  The theoretical framework presented here highlights the rich and complex dynamics that emerge with networked architectures.

\bibliography{ce_bib}

\end{document}